\documentclass[prb,twocolumn,showpacs,superscriptaddress,floatfix]{revtex4}
\usepackage{graphicx,amsfonts,amssymb,amsmath, hyperref}

\newif\ifhyper
\hypertrue
\ifhyper
\hypersetup{
   citecolor = {green},
   colorlinks = {true}, 
   urlcolor = {blue} 
}
\fi

\newcommand{\beq}{\begin{equation}}
\newcommand{\eeq}{\end{equation}}
\newcommand{\beqa}{\begin{eqnarray}}
\newcommand{\eeqa}{\end{eqnarray}}
\newcommand{\ket} [1] {\vert #1 \rangle}

\begin{document}

\title{Numerical study of the hard-core Bose-Hubbard model on an infinite square lattice}

\author{Jacob Jordan}
\email{jjordan@physics.uq.edu.au}
\affiliation{School of Physical Sciences, The University of Queensland, 
QLD 4072, Australia}

\author{Rom\'an Or\'us}
\email{orus@physics.uq.edu.au}
\affiliation{School of Physical Sciences, The University of Queensland, 
QLD 4072, Australia}

\author{Guifr\'e Vidal}
\email{vidal@physics.uq.edu.au}
\affiliation{School of Physical Sciences, The University of Queensland, 
QLD 4072, Australia}

\begin{abstract}
We present a study of the hard-core Bose-Hubbard model at zero temperature on an infinite square lattice using the infinite Projected Entangled Pair State algorithm [Jordan et al., Phys. Rev. Lett. 101, 250602 (2008)]. Throughout the whole phase diagram our values for the ground state energy, particle density and condensate fraction accurately reproduce those previously obtained by other methods. We also explore ground state entanglement, compute two-point correlators and conduct a fidelity-based analysis of the phase diagram. Furthermore, for illustrative purposes we simulate the response of the system when a perturbation is suddenly added to the Hamiltonian.
\end{abstract}
\maketitle

\section{Introduction}
 
The physics of interacting bosons at low temperature has since long attracted considerable interest due to the occurrence of Bose-Einstein condensation \cite{BEC}. The Bose-Hubbard model, a simplified microscopic description of an interacting boson gas in a lattice potential, is commonly used to study related phenomena, such as the superfluid-to-insulator transitions in liquid helium \cite{Fis89} or the onset of superconductivity in granular superconductors \cite{granular} and arrays of Josephson junctions \cite{Josephson}. In more recent years, the Bose-Hubbard model is also employed to describe  experiments with cold atoms trapped in optical lattices \cite{optical}.

As in most many-body systems, the theoretical study of interacting bosons cannot rely only on the few exact solutions available. Numerical results are also needed, but these are not always easy to obtain. Indeed, the exponential growth of the Hilbert space dimension in the lattice size (even after placing a bound on the number of bosons allowed on each of its sites) implies that exact diagonalization techniques are only capable of addressing very small lattices. Thus, in order to study the ground state properties of the Bose-Hubbard model on e.g. the square lattice, as is the goal of the present work, a number of more elaborate techniques, such as mean field theory, spin-wave calculations or quantum Monte Carlo are traditionally used (see e.g. \cite{Ber02} and references therein). 

Recently, a new class of simulation algorithms for two-dimensional systems, based on tensor networks, has gained much momentum. The basic idea is to use a network of tensors to efficiently represent the state of the lattice. Specifically, the so-called tensor product states\cite{TPS,TPS2} (TPS) or projected entangled-pair states\cite{PEPS,PEPSXX} (PEPS) are used to (approximately) express the $d^N$ coefficients of the wave function $\ket{\Psi}$ of a lattice of $N$ sites in terms of just $N$ tensors, in such a way that only $O(N)$ coefficients are actually specified. After optimizing these tensors so that $\ket{\Psi}$ represents e.g. the ground state of the system, one can then extract from them a number of properties, including the expected value of arbitrary local observables.
Moreover, in systems that are invariant under translations, the tensor network is made of copies of a small number of tensors. This leads to an even more compact description that depends on just $O(1)$ parameters. The later is the basis of the infinite PEPS (iPEPS) algorithm \cite{iPEPS}, which addresses infinite lattices and can thus be used to compute thermodynamic properties directly, without need to resort to finite size scaling techniques. 

In this work we initiate the exploration of interacting bosons in an infinite 2D lattice with tensor network algorithms. We use the iPEPS algorithm \cite{iPEPS,QCO} to characterize the ground state of the hard-core Bose-Hubbard (HCBH) model, namely the Bose Hubbard model in the hard-core limit, where either zero or one bosons are allowed on each lattice site. Although no analytical solution is known for the 2D HCBH model, there is already a wealth of numerical results based on mean-field theory, spin-wave corrections and stochastic series expansion \cite{Ber02}. These techniques have been quite successful in determining some of the properties of the ground state of the 2D HCHB model, such as its energy, particle density or condensate fraction. Our goal in this paper is twofold. Firstly, by comparing our results against those of Ref. \cite{Ber02}, we aim to benchmark the performance of the iPEPS algorithm in the HCBH model. Secondly, once the validity of the iPEPS algorithm for this model has been established, we use it to obtain results that are harder to compute with (or simply well beyond the reach of) the other approaches. These include the analysis of entanglement, two-point correlators, fidelities between different ground states\cite{Zan,Zho,Zho08}, and the simulation of time evolution.

We note that the present results naturally complement those of Ref. \cite{PEPSXX} for finite systems, where the PEPS algorithm \cite{PEPS} was used to study the HCBH model in a lattice made of at most $11\times 11$ sites. 

The rest of the paper is organized as follows. Sect. II introduces the HCBH model and briefly reviews the iPEPS algorithm. Sect. III contains our numerical results for the ground state of the 2D HCBH model. These include the computation of local observables such as the energy per lattice site, the particle density and the condensate fraction. We also analyze entanglement, two-point correlators and ground state fidelities. Finally, the simulation of time evolution is also considered. Sect. IV contains some conclusions. 

\section{Model and Method}

In this section we provide some basic background on the HCBH model, as well as on the iPEPS algorithm.

\subsection{The Hard Core Bose-Hubbard Model}

The Bose-Hubbard model \cite{Fis89} with on-site and nearest neighbour repulsion is described by the Hamiltonian
\beqa
H_{\mbox{\tiny BH}} = &-& J\sum\limits_{ \langle i,j \rangle } \left(a_i^\dag  a_j + a_j^\dag  a_i \right) 
~-~ \sum_i \mu \hat{n}_i  \nonumber \\
&+& \sum_iV_1 \hat{n}_i \left( \hat{n}_i  - 1 \right) ~+~ V_2 \sum_{ \langle i,j \rangle } \hat{n}_i \hat{n}_j  \nonumber \ ,
\eeqa
where $a_i^\dag$, $a_i$ are the usual bosonic creation and annihilation operators, $\hat{n}_i  = \hat{\rho}_i \equiv a_i^\dag  a_i^{}$ is the number (density) operator at site $i$, $J$ is the hopping strength, $\mu$ is the chemical potential, and $V_1, V_2 \ge 0$. The four terms in the above equation describe, respectively, the hopping of bosonic particles between adjacent sites ($J$), a single-site chemical potential ($\mu$), an on-site repulsive interaction ($V_1$) and an adjacent site repulsive interaction ($V_2)$. 

Here we shall restrict our attention to on-site repulsion only ($V_2$ = 0) and to the so-called \emph{hard-core limit} in which this on-site repulsion dominates ($V_1  \to \infty$). Under these conditions the local Hilbert space at every site describes the presence or absence of a single boson and has dimension 2. With the hard-core constraint in place, the Hamiltonian becomes
\beq
H_{\mbox{\tiny HC}}  =  - J\sum\limits_{ \langle i,j \rangle } \left(a_i^\dag a_j + a_j^\dag  a_i\right)  - \sum_i {\mu \hat{n}_i } \ , 
\label{hardcore}
\eeq
where $a_i^\dag$, $a_i$ are now hard-core bosonic operators obeying the commutation relation,
\beq
\left[ {a_i ,a_j^\dag } \right] = \left( {1 - 2\hat{n}_i} \right)\delta _{ij} \ . 
\nonumber
\eeq

A few well-known facts of the HCBH model are: 

(i) \emph{U(1) symmetry.---} The HCHB model inherits particle number conservation from the Bose-Hubbard model,
\begin{equation}
	[H_{\mbox{\tiny HC}},\hat{N}] = 0,~~~~~~~~~\hat{N} \equiv \sum_l \hat{n}_l \ ,
\end{equation}
and it thus has a $U(1)$ symmetry, corresponding to transforming each site $l$ by $e^{i\phi\hat{n}_l}$, $\phi \in [0,2\pi)$.

(ii) \emph{Duality transformation.---} In addition, the transformation $a_l \rightarrow a_l^{\dagger}$ applied on all sites $l$ of the lattice maps $H_{\mbox{\tiny HC}}(\mu)$ into $H_{\mbox{\tiny HC}}(-\mu)$ (up to an irrelevant additive constant). Accordingly, the model is self-dual at $\mu=0$, and results for, say, $\mu>0$ can be easily obtained from those for $\mu<0$.

(iii) \emph{Equivalence with a spin model.---} The HCBH model is equivalent to a quantum spin $\frac{1}{2}$ model, namely the ferromagnetic quantum XX model,
\begin{equation}
H_{\mbox{\tiny XX}}  =  - \frac{J}{2}\sum\limits_{ \langle i,j \rangle } {\sigma _i^x \sigma _j^x  + } \,\sigma _i^y \sigma _j^y  + \frac{\mu }{2}\sum\limits_i {\sigma _i^z} \ ,
\label{XX}
\end{equation}
which is obtained from $H_{\mbox{\tiny HC}}$ with the replacements
\beq
a_l  = \frac{{\sigma _l^x  + i\sigma _l^y }}{2} \ , ~~~ a^\dag_l  = \frac{{\sigma _l^x  - i\sigma _l^y }}{2} \ ,
\nonumber
\eeq
where $\sigma_x$, $\sigma_y$ and $\sigma_z$ are the spin $\frac{1}{2}$ Pauli matrices. In particular, all the results of this paper also apply, after a proper translation, to the ferromagnetic quantum XX model on an infinite square lattice.

(iv) \emph{Ground-state phase diagram.---} The hopping term in $H_{\mbox{\tiny HC}}$ favors delocalization of individual bosons in the ground state, whereas the chemical potential term determines the ground state bosonic \emph{density} $\rho$,
\beq
\rho \equiv \frac{1}{N}\sum\limits_{ i } \langle a_i^\dag  a_i\rangle  \ . 
\nonumber
\eeq
For $\mu$ negative, a sufficiently large value of $|\mu|$ forces the lattice to be completely empty, $\rho=0$. Similarly, a large value of (positive) $\mu$ forces the lattice to be completely full, $\rho=1$, as expected from the duality of the model. In both cases there is a gap in the energy spectrum and the system represents a Mott insulator. When, instead, the kinetic term dominates, the density has some intermediate value $0 < \rho < 1$, the cost of adding/removing bosons to the system vanishes, and the system is in a superfluid phase \cite{Fis89}. The latter is characterized by a finite fraction of bosons in the lowest momentum mode $\tilde{a}_{k=0} \equiv (1/N)\sum_{i} a_i$, that is by a non-vanishing \emph{condensate fraction} $\rho_0$,
\beq
\rho_0 \equiv \langle {\tilde{a}^\dag_{k = 0}\tilde{a}_{k = 0}} \rangle  = \frac{1}{N^2}\sum\limits_{ i,j } \langle a_j^\dag  a_i \rangle \ .
\nonumber
\eeq
In the thermodynamic limit, $N\rightarrow \infty$, a non-vanishing condensate fraction is only possible in the presence of off-diagonal long range order (ODLRO) \cite{ODLRO}, or $\langle a_j^\dag  a_i \rangle\neq 0$ in the limit of large distances $|i-j|$, given that
\begin{equation}
	\rho_0  = \lim_{|i-j| \rightarrow \infty}\langle a_j^\dag  a_i \rangle .
\end{equation}

(v) \emph{Quantum phase transition.---} Between the Mott insulator and superfluid phases, there is a continuous quantum phase transition \cite{Fis89}, tuned by \begin{math}\frac{\mu}{J}\end{math}. 


\begin{figure}
\includegraphics[width=7cm]{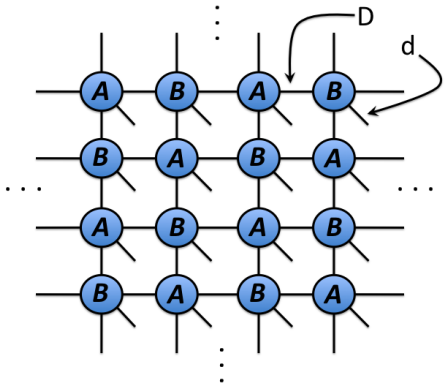}
\caption{(color online) Diagrammatic representation of a TPS/PEPS on a 2D square lattice. Tensors are represented by circles, and their indices are represented by legs. A leg connecting two circles corresponds to a \emph{bond index} shared by two tensors and takes $D$ different values. Since correlations between different sites of the lattice are carried by bond indices, the bond dimension $D$ is a measure of how many correlations the TPS/PEPS can represent. An open leg (diagonal line) corresponds to a \emph{physical index} that labels the local Hilbert space at a given lattice site. It takes $d$ different values, where $d$ is the local Hilbert space dimension (with $d=2$ for the HCBH model). Two different tensors, denoted $A$ and $B$, are repeated all over the infinite lattice, exploiting the fact that a translation invariant state is being represented. In principle, repeating a single tensor, say $A$, would be enough to represent a translation invariant state, but the iPEPS algorithm \cite{iPEPS} breaks translation invariance down to a checkerboard pattern.}
\label{fig:PEPS}
\end{figure}  


\subsection{The algorithm} 

The state $\ket{\Psi}$ of the infinite square lattice is represented using a TPS \cite{TPS} or PEPS \cite{PEPS} that consists of just two different tensors $A$ and $B$ that are repeated in a checkerboard pattern, see Fig. \ref{fig:PEPS}. Each of these two tensors depend on $O(dD^4)$ coefficients, where $d$ is the Hilbert space dimension of one lattice site (with $d=2$ for the HCBH model) and $D$ is a bond dimension that controls the amount of correlations or entanglement that the ansatz can carry. 

The coefficients of tensors $A$ and $B$ are determined with the iPEPS algorithm \cite{iPEPS}. Specifically, the ground state $\ket{\Psi_{\mbox{\tiny GS}}}$ of the HCBH model is obtained by simulating an evolution in imaginary time according to $H_{\mbox{\tiny HC}}$, exploiting that
\begin{equation}
	\ket{\Psi_{\mbox{\tiny GS}}} = \lim_{\tau\rightarrow \infty} \frac{e^{-\tau H_{\mbox{\tiny HC}}}\ket{\Psi_0}}{||e^{-\tau H_{\mbox{\tiny HC}}}\ket{\Psi_0}||}.
\end{equation}
We have also used the iPEPS algorithm to simulate (real) time evolution starting from the ground state $\ket{\Psi_{\mbox{\tiny GS}}}$ and according to a modified Hamiltonian $H$ (see Eq. \ref{eq:newH}),
\begin{equation}
	\ket{\Psi(t)} = e^{-it H}\ket{\Psi_{\mbox{\tiny GS}}}.
\end{equation}

These simulations, as well as the computation of expected values of local observables from the resulting state, involve contracting an infinite 2D tensor network. This is achieved with techniques developed for infinite 1D lattice systems \cite{iTEBD}, namely by evolving a matrix product state (MPS). An important parameter in these manipulations is the bond dimension $\chi$ of the MPS, which parameterizes how many correlations the latter can account for. We refer to \cite{iPEPS} for a detailed explanation of the iPEPS algorithm. In what follows we briefly comment on the main sources of errors and on the simulation costs.

We distinguish three main sources of errors in the simulations, one due to structural limitations in the underlying TPS/PEPS ansatz and two that originate in the particular way the iPEPS algorithm operates:

(i) \emph{Bond dimension $D$.---} A finite bond dimension $D$ limits the amount of correlations the TPS/PEPS can carry. A typical state of interest $\ket{\Psi}$, e.g. the ground state of a local Hamiltonian, requires in general a very large bond dimension $D$ if it is to be represented \emph{exactly}. However, a smaller value of $D$, say $D \geq D_{\Psi}$ for some value $D_{\Psi}$ that depends on $\ket{\Psi}$, often already leads to a good \emph{approximate} representation, in that the expected values of local observables are reproduced accurately. However, if $D < D_0$, then the numerical estimates may differ significantly from the exact values, indicating that the TPS/PEPS is not capable of accounting for all the correlations/entanglement in the target state $\ket{\Psi}$.

(ii) \emph{MPS bond dimension $\chi$.---} Similarly, using a finite MPS bond dimension $\chi$ implies that the contraction of the infinite 2D tensor network (required both in the simulation of real/imaginary time evolution and to compute expected values of local observables) is only approximate. This may introduce errors in the evolved state, or in the expected value of local observables even when the TPS/PEPS was an accurate representation of the intended state.

(iii) \emph{Time step.---} A time evolution (both in real or imaginary time) is simulated by using a Suzuki-Trotter expansion of the evolution operator ($e^{-i t H}$ or $e^{-\tau H}$), which involves a time step ($\delta t$ or $\delta \tau$). This time step introduces an error in the evolution that scales as some power of the time step. Therefore this error can be reduced by simply diminishing the time step. 

The cost of the simulations scales as $O(\chi^3D^6+\chi^2D^8d)$ (here we indicate only the leading orders in $\chi$ and $D$; the cost of the simulation is also roughly proportional to the inverse of the time step). This scaling implies that only small values of the bond dimensions $D$ and $\chi$ can be used in practice. In our simulations, given a value of $D$ ($D=2,3$ or $4$), we choose a sufficiently large $\chi$ (in the range $10-40$) and sufficiently small time step ($\delta t$ or $\delta \tau$) such that the results no longer depend significantly on these two parameters. In this way the bond dimension $D$ is the only parameter on which the accuracy of our results depends.

On a 2.4 GHz dual core desktop with 4 Gb of RAM, computing a superfluid ground state (e.g. $\mu=0$) with $D=2$, $\chi=20$ and with $\delta \tau$ decreasing from $10^{-1}$ to $10^{-4}$ requires about 12 hours. Computing the same ground state with $D=3$ and $\chi=40$ takes of the order of two weeks.

\section{Results}

In this section we present the numerical results obtained with the iPEPS algorithm. 

Without loss of generality, we fix the hopping strength $J=1$ and compute an approximation to the ground state $\ket{\Psi_{\mbox{\tiny GS}}}$ of $H_{\mbox{\tiny HC}}$ for different values of the chemical potential $\mu$. Then we use the resulting TPS/PEPS to extract the expected value of local observables, analyze ground state entanglement, compute two-point correlators and fidelities, or as the starting point for an evolution in real time. 

In most cases we only report results for $\mu \leq 0$ (equivalently, density $0\leq \rho \leq 0.5$) since due to the duality of the model, results for positive $\mu$ (equivalently, $0.5 \leq \rho \leq 1$) can be obtained from those for negative $\mu$.

\subsection{Local observables and phase diagram}

\begin{figure}
\includegraphics[width=9cm]{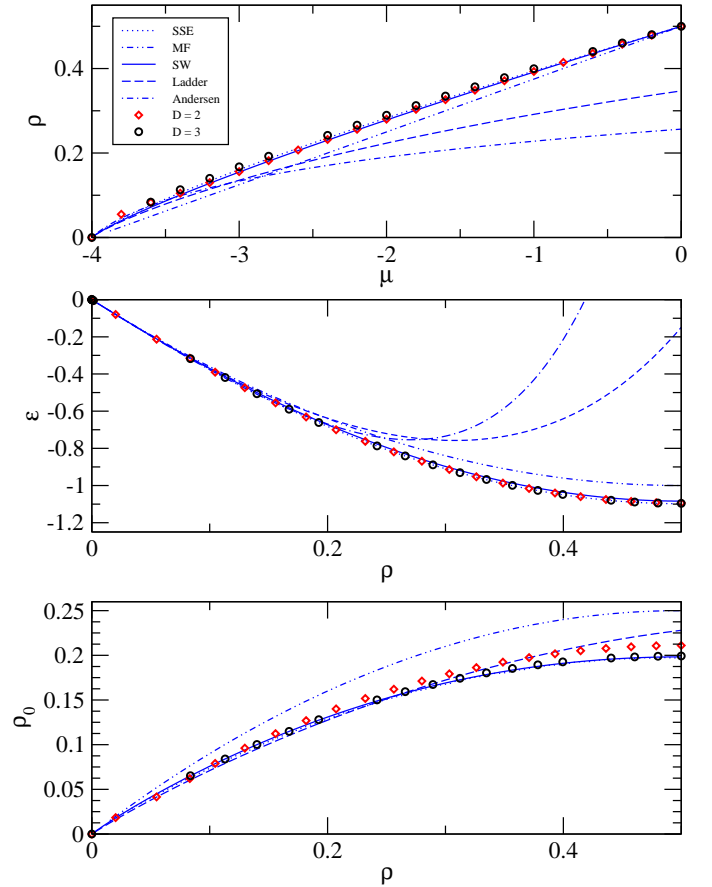}
\caption{(color online) Particle density $\rho(\mu)$, energy per lattice site $\epsilon(\rho)$ and condensate fraction $\rho_0(\rho)$ for a TPS/PEPS with $D = 2,3$. We have also plotted results from Ref.\cite{Ber02} corresponding to several other techniques. Our results follow closely those obtained with stochastic series expansion (SSE) and mean field with spin wave corrections (SW).}
\label{fig:LO}
\end{figure}

\emph{Particle density $\rho$.---} Fig. \ref{fig:LO} shows the density $\rho$ as a function of the chemical potential $\mu$ in the interval $-4 \leq \mu \leq 0$. Notice that $\rho=0$ for $\mu \leq -4$, since each single site is vacant. Our results are in remarkable agreement with those obtained in Ref. \cite{Ber02} with stochastic series expansions (SSE) for a finite lattice made of $32\times 32$ and with a mean field calculations plus spin wave corrections (SW). We note that the curves $\rho(\mu)$ for $D=2$ and $D=3$ are very similar.

\emph{Energy per site $\epsilon$.---} Fig. \ref{fig:LO} also shows the energy per site $\epsilon$ as a function of the density $\rho$. This is obtained by computing $\epsilon(\mu)$ and then replacing the dependence on $\mu$ with $\rho$ by inverting the curve $\rho(\mu)$ discussed above. Again, our results for $\epsilon(\rho)$ are in remarkable agreement with those obtained in Ref. \cite{Ber02} with stochastic series expansions (SSE) for a finite lattice made of $32\times 32$. They are also very similar to the results coming from mean field calculations with spin wave corrections (SW) of Ref \cite{Ber02}, and for small densities reproduce the scaling (valid only in the regime of a very dilute gas) predicted in Ref. \cite{Hines} by using field theory methods based on a summation of ladder diagrams. Once more, the curves $\epsilon(\rho)$ obtained with bond dimension $D=2$ and $D=3$ are very similar, although $D=3$ produces slightly lower energies.

\emph{Condensate fraction $\rho_0$.---} In order to compute the condensate fraction $\rho_0$, we exploit that the iPEPS algorithm induces a spontaneous symmetry breaking of particle number conservation. Indeed, one of the effects of having a finite bond dimension $D$ is that the TPS/PEPS that minimizes the energy does not have a well-defined particle number. As a result, instead of having $\langle a_i \rangle = 0$, we obtain a non-vanishing value $\langle a_i \rangle \neq 0$ such that
\begin{equation}
	\rho_0 = \lim_{|i-j|\rightarrow \infty} \langle a_j^\dag  a_i \rangle = |\langle a_i \rangle|^2.
\end{equation}
In other words, the ODLRO associated with the presence of superfluidity, or a finite condensate fraction, can be computed by analysing the expected value of $a_l$,
\begin{equation}
	\langle a_l \rangle = \sqrt{\rho_0}e^{i\varphi},
	\label{eq:phase}
\end{equation}
where the phase $\varphi$ is constant over the whole system but is otherwise arbitrary. The condensate fraction $\rho_0$ shows that the model is in an insulating phase for $|\mu| \geq 4$ ($\rho = 0,1$) and in a superfluid phase for $-4 < \mu < 4$ ($0 < \rho < 1$), with a continuous quantum phase transition occurring at $|\mu| = -4$, as expected. However, this time the curves $\rho_{0}(\rho)$ obtained with $D=2$ and $D=3$ are noticeably different, with $D=3$ results again in remarkable agreement with the SSE and SW results of Ref. \cite{Ber02}.

\subsection{Entanglement}

The iPEPS algorithm is based on assuming that a TPS/PEPS offers a good description of the state $\ket{\Psi}$ of the system. Results for small $D$ will only be reliable if $\ket{\Psi}$ has at most a moderate amount of entanglement. Thus, in order to understand in which regime the iPEPS algorithm should be expected to provide reliable results, it is worth studying how entangled the ground state $\ket{\Psi_{\mbox{\tiny GS}}}$ is as a function of $\mu$. 

The entanglement between one site and the rest of the lattice can be measured by the degree of purity of the reduced density matrix $\varrho_1$ for that site,
\beq
\varrho_1 = \frac{\mathbb{I} + \vec{r} \cdot \vec{\sigma}}{2} \ ,~~~~~\vec{\sigma}\equiv(\sigma_x,\sigma_y,\sigma_z),
\eeq
as given by the norm $r$ of the Bloch vector $\vec{r}$. If the lattice is in a product or unentangled state, then each site is in a pure state, corresponding to purity $r=1$. On the other hand, if the lattice is in an entangled state, then the one-site reduced density matrix will be mixed, corresponding to purity $r < 1$. Accordingly, one can think of $r$ as measuring the amount of entanglement between one site and the rest of the lattice, with less purity corresponding to more entanglement. 


\begin{figure}
\includegraphics[width=8.5cm]{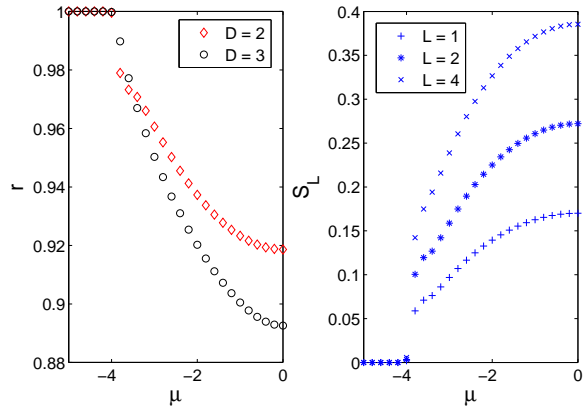}
\caption{(Color online) Purity $r$ and entanglement entropy $S_L$ as a function of the chemical potential $\mu$. The results indicate that the ground state is more entangled deep inside the superfluid phase ($\mu=0$) than at the phase transition point ($\mu=-4$). Notice that the more entangled the ground state is, the larger the differences between results obtained with $D=2$ and $D=3$ (see also Fig. \ref{fig:LO}).}
\label{fig:entanglement}
\end{figure}

Fig. \ref{fig:entanglement} shows the purity $r$ as a function of the chemical potential. In the insulating phase ($\mu \leq -4$), the ground state of the system consists of a vacancy on each site. In other words, it is a product state, $r=1$. Instead, For $\mu > -4$ the ground state is entangled. Several comments are in order:

(i) The purity $r(\mu)$ for $D=3$ is smaller than that for $D=2$ by up to $3\%$. This is compatible with the fact that the a TPS/PEPS with larger bond dimension $D$ can carry more entanglement. 

(ii) Results for $D=2,3$ seem to indicate that the ground state is more entangled ($r$ is smaller) deep into the superfluid phase (e.g. $\mu = 0$) than at the continuous quantum phase transition $\mu=-4$. This is in sharp contrast with the results obtained e.g. for the 2D quantum Ising model \cite{iPEPS}, where the quantum phase transition displays the most entangled ground state. However, notice that in the Ising model the system is only critical at the phase transition whereas in the present case criticality extends throughout the superfluid phase. Each value of $\mu$ in the superfluid phase corresponds to a fixed point of the RG flow. That is, in moving away from the phase transition we are not following an RG flow. Therefore, the notion that entanglement should decrease along an RG flow\cite{RGFlow}, as observed in the 2D Ising model, is not applicable for the HCBH model. 

(iii) Accordingly, we expect that the iPEPS results for small $D$ become less accurate as we go deeper into the superfluid phase (that is, as we approach $\rho = 0.5$). This is precisely what we observe: the curves $\rho_0(\rho)$ for $D=2$ and $D=3$ in Fig. \ref{fig:LO} differ most at $\rho=0.5$.

Fig. \ref{fig:entanglement} also shows the entanglement entropy
\begin{equation}
	S(\varrho_L) \equiv - tr(\varrho_L \ln \varrho_L )
\end{equation}
for the reduced density matrix $\varrho_L$ ($L=1,2,4$) corresponding to one site, two contiguous sites and a block of $2\times 2$ sites respectively. The entanglement entropy vanishes for an unentangled state and is non-zero for an entangled state. The curves $S(\varrho_L)$ confirm that the ground state of the HCBH model is more entangled deep in the superfluid phase than at the quantum phase transition point.

\subsection{Correlations}

\begin{figure}
\includegraphics[width=9cm]{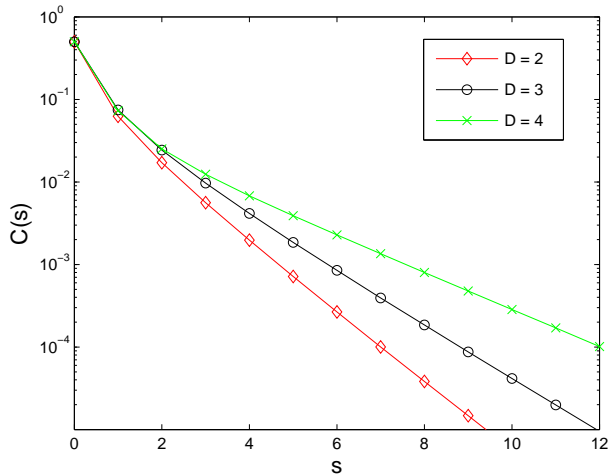}
\caption{(Color online) Two-point correlation function $C(s)$ versus distance $s$ (measured in lattice sites), along a horizontal direction of the lattice. For very short distances the correlator for $D=2,3,4$ are very similar whereas for larger distances they differ significantly.}
\label{fig:corr}
\end{figure}

From a TPS/PEPS for the ground state $\ket{\Psi_{\mbox{\tiny GS}}}$ it is easy to extract equal-time two-point correlators. For illustrative purposes, Fig. (\ref{fig:corr}) shows a connected two-point correlation function $C(s)$, 
\beq
C(s) \equiv \langle a^\dag_{i} a_{i+s\hat{x}} \rangle - \langle a^\dag_{i}\rangle  \langle a_{[i+s\hat{x}]} \rangle\ ,
\eeq
between two sites that separated $s$ lattice sites along the horizontal direction $\hat{x}$. The plot corresponds to a superfluid ground state, $\mu=0$, where $C(s)$ displays an exponential decay, in spite of the fact that the Hamiltonian is gapless.

The results show that while for short distances $s=0,1,2$ the correlator $C(s)$ is already well converged with respect to $D$, for larger distances $s$ the correlator still depends significantly on $D$. This seems to indicate that while the iPEPS algorithm provides remarkably good results for local observables already for affordably small values of $D$, a larger $D$ might be required in order to also obtain accurate estimates for distant correlators.

\subsection{Fidelity}

Given two ground states $\ket{\Psi_{\mbox{\tiny GS}}(\mu_1)}$ and $\ket{\Psi_{\mbox{\tiny GS}}(\mu_2)}$, corresponding to different chemical potential $\mu$, the fidelity per site $f$ \cite{Zho}, defined through
\beq
\ln f (\mu_1 ,\mu_2 ) = \mathop {\lim }\limits_{N \to \infty } \frac{1}{N}\ln | \langle \Psi_{\mbox{\tiny GS}}(\mu_1) | \Psi_{\mbox{\tiny GS}}(\mu_2)\rangle | \ ,
\nonumber
\eeq
can be used as a means to distinguish between qualitatively different ground states \cite{Zan,Zho}. In the above expression, $N$ is the number of lattice sites and the thermodynamic limit $N \to \infty$ is taken. Importantly, the fidelity per site $f(\mu_1,\mu_2)$ remains finite in this limit, even though the overall fidelity $|\langle \Psi_{\mbox{\tiny GS}}(\mu_1) | \Psi_{\mbox{\tiny GS}}(\mu_2)\rangle |$ vanishes. In a sense, $f(\mu_1,\mu_2)$ captures how quickly the overall fidelity vanishes.

\begin{figure}
\includegraphics[width=8cm]{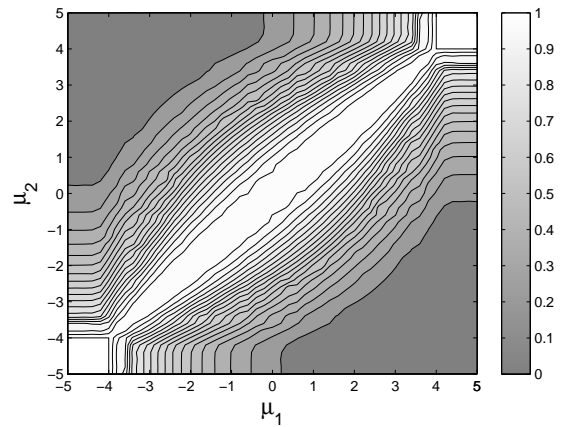}
\caption{Fidelity per lattice site $f(\mu_1,\mu_2)$ for the ground states of the HCBH model. Notice the plateau $f(\mu_1,\mu_2)=1$ (white) for $\mu_1,\mu_2 \leq -4$ (also for $\mu_1,\mu_2 \geq 4$) corresponding to the Mott insulating phase, and the pinch point at $\mu_1,\mu_2 = -4$  (also at $\mu_1,\mu_2 = 4$) consistent with a continuous quantum phase transition.}
\label{fig:fidelity}
\end{figure} 

Fortunately, the fidelity per site $f(\mu_1,\mu_2)$ can be easily computed within the framework of the iPEPS algorithm \cite{Zho08}. In the present case, before computing the overlap each ground state is rotated according to $e^{i\varphi\sigma_z/2}$, where $\varphi$ is the random condensate phase of Eq. \ref{eq:phase}. In this way all the ground states have the same phase $\varphi=0$. The fidelity per site $f(\mu_1,\mu_2)$ is presented in Fig. \ref{fig:fidelity}. 
The plateau-like behavior of $f(\mu_1,\mu_2)$ for points within the separable Mott-Insulator phase ($\mu_1,\mu_2 \leq -4$ or $\mu_1,\mu_2 \geq 4$) is markedly different from that between ground states in the superfluid region ($-4 \leq \mu_1,\mu_2 \leq 4$), where the properties of the system vary continuously. Moreover, similarly to what has been observed for the 2D quantum Ising model \cite{Zho08} or in the 2D quantum XYX model \cite{NewZho08}, the presence of a continuous quantum phase transition between insulating and superfluid phases in the 2D HCBH model is signaled by \emph{pinch points} of $f(\mu_1,\mu_2)$ at $\mu_1=\mu_2 = \pm 4$. That is, the qualitative change in ground state properties across the critical point is evidenced by a rapid, continuous change in the fidelity per lattice site as one considers two ground states on opposite sides of the critical point and moves away from it. 

\subsection{Time evolution}

An attractive feature of the algorithms based on tensor networks is the possibility to simulate (real) time evolution. A first example of such simulations with the iPEPS algorithm was provided in Ref. \cite{QCO}, where an adiabatic evolution across the quantum phase transition of the 2D quantum compass orbital model was simulated in order to show that the transition is of first order.

The main difficulty in simulating a (real) time evolution is that, even when the initial state $\ket{\Psi(0)}$ is not very entangled and therefore can be properly represented with a TPS/PEPS with small bond dimension $D$, entanglement in the evolved state $\ket{\Psi(t)}$ will typically grow with time $t$ and a small $D$ will quickly become insufficient. Incrementing $D$ results in a huge increment in computational costs, which means that only those rare evolutions where no much entanglement is created can be simulated in practice.

\begin{figure}
\includegraphics[width=8cm]{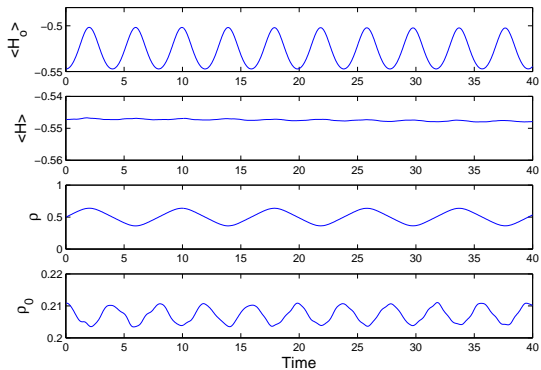}
\caption{(Color online) Evolution of the energies $\langle H_0\rangle$ and $\langle H \rangle$, the density $\rho$, and condensate fraction $\rho_0$ after a translation invariant perturbation $V$ is suddenly added to the Hamiltonian. }
\label{fig:TE}
\end{figure} 

For demonstrative purposes, here we have simulated the response of the ground state $\ket{\Psi_{\mbox{\tiny GS}}}$ of the HCBH model at half filling ($\rho=0.5$ or $\mu=0$) when the Hamiltonian $H_{\mbox{\tiny HC}}$ is suddenly replaced with a new Hamiltonian $H$ given by
\beq
H  \equiv H_{\mbox{\tiny HC}} + \gamma V,~~~~~~~~V\equiv   - i \sum_k \left({a_k - a_k^\dag}\right) \ , 
\label{eq:newH}
\eeq
where $\gamma=0.2$ and, importantly, the perturbation $V$ respects translation invariance. As the starting point of the simulation, we consider a TPS/PEPS representation of the ground state with bond dimension $D=2$, obtained as before through imaginary time evolution.

Fig. \ref{fig:TE} shows the evolution in time of the expected value per site of the energies $\langle H_{\mbox{\tiny HC}}\rangle$ and $\left\langle {H } \right\rangle$, as well as the density $\rho$ and condensate fraction $\rho_0$. Notice that the expected value of $H$ should remain constant through the evolution. The fluctuations observed in $\left\langle {H } \right\rangle$, of the order of 0.2\% of its total value, are likely to be due to the small bond dimension $D=2$ and indicate the scale of the error in the evolution. The simulation shows that, as a result of having introduced a perturbation $V$ that does not preserve particle number, the particle density $\rho$ oscillates in time. The condensate fraction, as measured by $|\langle a_l\rangle|^2$, is seen to oscillate twice as fast.

\section{Conclusion}

In this paper we have initiated the study of interacting bosons on an infinite 2D lattice using the iPEPS algorithm. We have computed the ground state of the HCBH model on the square lattice as a function of the chemical potential. Then we have studied a number of properties, including properties that can be easily accessed with other techniques \cite{Ber02}, as is the case of the expected value of local observables, as well as properties whose computation is harder, or even not possible, with previous techniques.

Specifically, using a small bond dimension $D=2,3$ we have been able to accurately reproduce the result of previous computations using SSE and SW of Ref. \cite{Ber02} for the expected value of the particle density $\rho$, energy per particle $\epsilon$ and condensate fraction $\rho_0$, throughout the whole phase diagram of the model, which includes both a Mott insulating phase and a superfluid phase, as well as a continuous phase transition between them. Interestingly, in the superfluid phase the TPS/PEPS representation spontaneously breaks particle number conservation, and the condensate fraction can be computed from the expected value of the annihilation operator, $\rho_0 = |\langle a_l \rangle|^2$.

We have also conducted an analysis of entanglement, which revealed that the most entangled ground state corresponds to half filling, $\rho=0.5$. This is deep into the superfluid phase and not near the phase transition, as in the case of the 2D quantum Ising model\cite{iPEPS}. Furthermore, inspection of a two-point correlator at half filling showed much faster convergence in the bond dimension $D$ for short distances than for large distances. Also, pinch points in plot of the fidelity $f(\mu_1,\mu_2)$ were consistent with continuous quantum phase transitions at $\mu = \pm 4$.

Finally, we have also simulated the evolution of the system, initially in the ground state of the HCHB model at half filling, when a translation invariant perturbation is suddenly added to the Hamiltonian. 

Now that the validity of the iPEPS algorithm for the HCBH model (equivalently, the quantum XX spin model) has been established, there are many directions in which the present work can be extended. For instance, one can easily include nearest neighbor repulsion, $V_2 \ne 0$, (corresponding to the quantum XXZ spin model) and/or investigate a softer-core version of the Bose Hubbard model by allowing up to two or three particles per site.

{\it Acknowledgements.-} We thank I. McCulloch and M. Troyer for stimulating discussions, and G. Batrouni for the generous provision of comparitive data. Support from The University of Queensland (ECR2007002059) and the Australian Research Council (FF0668731, DP0878830) is acknowledged.

\end{document}